\documentclass[twocolumn,preprintnumbers,showpacs,pre,floatfix,superscriptaddress,amsmath,amssymb]{revtex4-1}
\usepackage{epsfig}
\usepackage{subfigure}
\usepackage{amsmath}
\usepackage{color}
\usepackage{amssymb}
\usepackage{setspace}
\usepackage{graphicx}
\usepackage{dcolumn}
\usepackage{bm}
\usepackage{times}
\input epsf

\begin{document}

\author{Dane Taylor}
\email{dane.taylor@colorado.edu} 
\affiliation{Department of Electrical, Computer, and Energy Engineering, University of Colorado, Boulder, Colorado 80309,
USA}

\author{Edward Ott}
\affiliation{Institute for Research in Electronics and Applied Physics, University of Maryland, College Park, Maryland 20742,
USA}

\author{Juan G. Restrepo}
\affiliation{Applied Mathematics Department, University of Colorado, Boulder, Colorado 80309,
USA}

\date{\today}

\begin{abstract}
We study the synchronization of Kuramoto oscillators with all-to-all coupling in the presence of slow, noisy frequency adaptation. In this paper we develop a new model for oscillators which adapt both their phases and frequencies. It is found that this model naturally reproduces some observed phenomena that are not qualitatively produced by the standard Kuramoto model, such as long waiting times before the synchronization of clapping audiences.  By assuming a self-consistent steady state solution, we find three stability regimes for the coupling constant $k$, separated by critical points $k_1$ and $k_2$: (i) for $k<k_1$ only the stable incoherent state exists; (ii) for $k>k_2$, the incoherent state becomes unstable and only the synchronized state exists; (iii) for $k_1<k<k_2$ both the incoherent and synchronized states are stable. In the bistable regime spontaneous transitions between the incoherent and synchronized states are observed for finite ensembles. These transitions are well described as a stochastic process on the order parameter $r$ undergoing fluctuations due to the system's finite size, leading to the following conclusions: (a) in the bistable regime, the average waiting time of an incoherent$\rightarrow$coherent transition can be predicted by using Kramer's escape time formula and grows exponentially with the number of oscillators; (b) when the incoherent state is unstable ($k>k_2$), the average waiting time grows logarithmically with the number of oscillators.
\end{abstract}

\title{Spontaneous synchronization of coupled oscillator systems with frequency adaptation}
\maketitle

\section{Introduction}\label{introd}
Many natural and engineered systems can be described as an ensemble of  heterogeneous limit-cycle oscillators influencing each other. Examples include glycolytic oscillations in yeast cell populations \cite{dano}, pedestrians walking over a bridge \cite{millenium}, arrays of Josephson junctions \cite{josephson}, the power grid \cite{Fillatrella}, lasers \cite{Cross}, and some species of fireflies \cite{ermentrout}. A central issue is that of understanding the mechanism of coherent behavior that is often observed for these systems.

The Kuramoto model \cite{kuramoto} (for a review, see \cite{kuramotoreview}) addresses this problem by considering the simplified case in which oscillators 
are all-to-all coupled, and each oscillator, labeled by an index $n$, has an intrinsic frequency $\omega_n$ and an oscillator state that can be specified solely by its phase angle $\theta_n$. The evolution of the phase of each oscillator $n$ is given by
\begin{equation}\label{kuramotoeq}
\dot{\theta}_n = \omega_n + \frac{k}{N} \sum_{m=1}^N\sin(\theta_m-\theta_n),
\end{equation}
where $N$ is the number of oscillators, $m = 1,2,\dots,N$ indicates different oscillators, and $k$ is a parameter that represents the strength of the coupling between oscillators. Kuramoto found that for the $N\to\infty$ coupling limit (approximating the typical case of large $N$ arising in most applications), the collective behavior of the oscillator ensemble, quantified by the order parameter $r = |\sum_{m=1}^N \exp(i\theta_m)|$, undergoes a transition from incoherence ($r = 0$) to synchronization ($r\sim 1$) as the coupling strength is increased past a critical value $k_c$. The Kuramoto model provides a simple mathematical model capturing the essential mechanisms for synchronization of limit-cycle oscillators. Despite its long-standing status as a classical model of synchronization, some advances in the theoretical understanding of Kuramoto-type models have been achieved only very recently (e.g., Refs.~\cite{otts}).

Due to the ubiquity of synchronization phenomena in complex systems, there is current interest in understanding the effect of network structure interactions and adaptation on the synchronization of oscillators \cite{adapta}.  The new model presented in this Article aims to investigate synchronization in coupled oscillator systems where each oscillator's natural frequency $\omega_n$ slowly adapts, while being subjected to random noise-like fluctuations. One motivation for considering adaptive synchronization is the observation that, in many biological situations, synchronization seems to serve a useful function. A fairly clear example is the synchronization of pacemaker cells in the heart. Another example is the observed evolving patterns of neuronal synchrony in the brain, which have been conjectured to play a key role in organizing brain function. In some cases, adaption of frequencies has been experimentally observed: fireflies of the species pteroptyx-malaccae slowly adapt their flashing frequency in response to the flashes they observe \cite{ermentrout}. Assuming the utility of synchronization in such biological cases, it is reasonable that there might be evolutionary pressure for the development of adaptive mechanisms that promote synchronization or maintain it in the presence of disruptive influences (e.g., noise). In addition, one could imagine technological and social situations where adaptation to promote synchronization might be relevant. A familiar social example is that of an audience clapping their hands and seeking to synchronize \cite{clapping,Xenides}. Therefore, it is important to consider the possibility that various mechanisms might operate independently to promote synchronization in a noisy environment. In this paper we introduce and analyze a model of all-to-all coupled phase oscillators with noisy frequency adaptation, where, as seems reasonable in the above-cited biological examples, the coupling of the oscillators' phases  occurs on a faster timescale than the frequency adaptation dynamics.

The paper is organized as follows. In Sec.~\ref{secmodel} we present and analyze our model. Analytical results are derived and numerically tested  in Secs.~\ref{secmodela} and \ref{secmodelb}, respectively. In Sec.~\ref{transitions} we study the statistics of spontaneous transitions to the synchronized state due to finite size effects. In Sec.~\ref{secdiscussion} we discuss our results and their relation with previous work. In Sec.~\ref{secconclusions} we present our conclusions.

\section{Frequency Adaption model}\label{secmodel}
We consider the classical Kuramoto model supplemented with a dynamical equation for the evolution of the oscillators' natural frequencies $\omega_n$,
\begin{align}\label{model}
\dot{\theta}_n &= \omega_n + \frac{k}{N} \sum_{m=1}^N\sin(\theta_m-\theta_n),\\
\dot{\omega}_n &= \tau^{-1} \frac{k}{N} \sum_{m=1}^N\sin(\theta_m-\theta_n) +\eta_n\nonumber -dV_n(\omega_n)/d\omega_n,
\end{align}
where $\tau$ is assumed to be much larger than the spread in the oscillator period and $\eta_n$ is a Gaussian uncorrelated noise term such that $\langle \eta_n(t)\eta_m(t')\rangle = 2D\delta_{nm}\delta(t-t')$, where $\langle\cdot\rangle$ is an ensemble average and $\delta_{nm}$ is the Kronecker delta. The motivation for the different terms in this natural frequency adaption term is the following:
\begin{itemize}
 \item We assume that each oscillator $n$ may only have knowledge of the aggregate input to it from the other oscillators, $k \sum_{m=1}^N\sin(\theta_m-\theta_n)$. This frequency-coupling term was originally introduced in Ref.~\cite{ermentrout}, who considered frequency adaptation without phase coupling. 
  \item The form of the coupling guarantees that if the phase of oscillator $n$ is behind (ahead of) the average phase [so that $\sin(\theta_m-\theta_n)$ is, on average, positive (negative)], its frequency increases (decreases).
   \item Frequency adaptation occurs on a time scale $\tau$, much slower than the phase dynamics.
   \item The intrinsic frequencies $\omega_n$ are subject to random noise $\eta_n$. This is partly motivated by observations of frequency drift in biological oscillators \cite{drift}. 
   \item The confining potential $V_n(\omega)$ represents physical mechanisms that, depending on the application, constrain the natural frequencies to some reasonable range. 
\end{itemize}

The dynamics we find for our system ({\ref{model}) is related to that for the Kuramoto model with inertia {\cite{inertia1,inertia2}; however, the differences are significant and will be discussed in Sec.~\ref{secdiscussion}. Also in Sec.~\ref{secdiscussion}, we discuss the relation between ({\ref{model}) and a model for circadian rhythms that implements wandering, uncoupled frequencies \cite{Rougemont}.

We note that we could have added a noise term to the $\theta$ equation in (\ref{model}). However, the effect of such a noise term has been already studied in the Kuramoto model, and it has been found that it shifts the transition to synchronization to larger values of the coupling strength $k$, maintaining the same qualitative behavior. Therefore, for analytical simplicity, we will not consider this term. As we will see in our numerical simulations, a role analogous to fluctuations in $\theta$ will be played by the fluctuations resulting from having a finite number of oscillators.

\subsection{Model Analysis}\label{secmodela}
We consider $N$ to be very large and adopt a continuum description. Thus, we assume that the ensemble of oscillator intrinsic frequencies can be regarded as being drawn from a continuous distribution function $G(\omega,t)$.

We analyze our proposed model (\ref{model}) by using the assumed separation of timescales between the oscillator dynamics and the frequency adaptation. Rewritting Eqs.~(\ref{model}) in terms of the mean field $r e^{i\psi} =\frac{1}{N} \sum_{m=1}^Ne^{i \theta_m}$, we obtain
\begin{align}
\dot{\theta}_n &= \omega_n - k r \sin(\theta_n -\psi),\label{kusta}\\
\dot{\omega}_n &= -\tau^{-1} k r \sin(\theta_n -\psi) + \eta_n - dV(\omega_n)/d\omega_n.\label{omesta}
\end{align}
Here we have dropped the subscript $n$ on the potential $V_n$, as we will henceforth consider all oscillators to have the same $V_n$. In addition, we will assume that $V(\omega) = V(-\omega)$.

Since the frequencies vary on a timescale much longer than the phases, on the fast time scale we can approximate $\omega_n$ in Eq.~(\ref{kusta}) as constant.  
As we shall soon see, it is relevant to assume that $G(\omega,t)$ is symmetric in $\omega$ and monotonically decreasing away from its maximum. Furthermore, without loss of generality, it suffices to take the maximum of $G$ to occur at $\omega =0$ (if it does not, it can be shifted to $0$ by the change of variables $\omega\to \omega +\Omega$, $\theta\to \theta  - \Omega t$). As originally noted by Kuramoto, in the saturated state (i.e., $r$ constant on the fast time scale), the phase dynamics is of two types depending on the value of $\omega_n$. For $|\omega_n|< kr$ oscillator $n$ is said to be ``locked'' and its phase settles at a value given by
\begin{equation}\label{eq5}
\sin(\theta_n-\psi) = \omega_n/(kr).
\end{equation}
For $|\omega_n| > kr$ the phase is said to ``drift'' and $\theta_n$ continually increases (decreases) with time for $\omega_n > kr$ ($\omega_n <- kr$).
For a given frequency $|\omega| > kr$, the drifting oscillators have a distribution of phases $\rho(\theta,\omega)$ determined from the conservation of oscillator density by the condition $\rho d\theta/dt = $ constant. This yields
\begin{equation}\label{ave1}
\rho(\theta,\omega) = \frac{\sqrt{\omega^2-(kr)^2}}{2\pi|\omega - kr \sin(\theta-\psi)|},
\end{equation}
where the factor $\sqrt{\omega^2 - (kr)^2} /(2\pi)$ normalizes $\rho(\theta,\omega)$ so that $\int_{-\pi}^{\pi}\rho d\theta = 1$.

Still invoking the time scale separation, and consequently assuming that the deterministic term in Eq.~(\ref{omesta}) can be averaged over time, we obtain an approximation to Eq.~(\ref{omesta}) for the drifting oscillators
\begin{equation}\label{w2}
\dot \omega_n \approx -\tau^{-1} kr \int_{-\pi}^{\pi}\rho(\theta,\omega_n)\sin(\theta-\psi)d\theta + \eta_n -dV/d\omega_n.
\end{equation}
[Here the deterministic term has been replaced by its time average, while the fast-varying noise term has been retained. This can be justified by noting that the difference between the original equation, Eq.~(\ref{omesta}), and the equation where the deterministic part was averaged, Eq.~(\ref{w2}), is what would be obtained in the noiseless case. Since in the noiseless case this difference can be argued to be small by averaging, we conclude our procedure is justified.]

Integrating Eq.~(\ref{w2}) and recalling that for entrained oscillators $kr\sin(\theta_n-\psi) = \omega_n$, Eq.~(\ref{omesta}) can be rewritten as
\begin{equation}\label{drift}
\dot \omega_n = h(\omega_n) +\eta_n -dV/d\omega_n,
\end{equation}
where
\begin{align}\label{sode}
h(\omega) = \left\{ \begin{array}{ll}
         -\omega/\tau ,& |\omega| \leq k r,\nonumber\\
         -\omega/\tau +\mbox{sign}(\omega)\sqrt{\omega^2-(kr)^2}/\tau,& |\omega| > k r.\end{array} \right. 
\end{align}

We now seek a steady state solution (in a statistical sense) for Eqs.~(\ref{model}). More precisely, for a given value of $k$, we seek a time-independent probability distribution of frequencies $G$ and a value of $r$ that make Eqs.~(\ref{model}) consistent. Such a steady-state frequency distribution  can be obtained by solving the time-independent Fokker-Planck equation corresponding to Eq.~(\ref{drift})
\begin{equation}\label{fp}
\frac{d}{d\omega} \left[\left(h-\frac{dV}{d\omega}\right)G\right]=D\frac{d^2 G}{d\omega^2}.
\end{equation}\\

The solution of Eq.~(\ref{fp}) with no-flux boundary conditions (${dG}/{d\omega}=0$ at either $\omega = \pm \infty$ or $\omega = \pm L$) is
$
G \propto \exp[\int h(\omega)d\omega/D - V(\omega)/D],
$
from which we obtain Eq.~(\ref{ge}), where $\sigma^2 = D\tau$:
\begin{widetext}
\begin{align}\label{ge}
G(\omega,kr) \propto \left\{ \begin{array}{ll}
	\exp[-\omega^2/(2\sigma^2) -V(\omega)/\sigma^2],&  |\omega| \leq k r,\\
	\left[\frac{|\omega|}{kr}\left(1-\sqrt{1-(kr/\omega)^2}\right)\right]^{(kr)^2/(2\sigma^2)}\exp\left[-\frac{\omega^2}{2\sigma^2}\left(1-\sqrt{1-(kr/\omega)^2}\right)-V(\omega)/\sigma^2\right] ,&  |\omega| > k r.\\
\end{array} \right. 
\end{align}
\end{widetext}

In all our numerical plots we use $D = 0.01$ and $\tau = 50$, which yields $\sigma^2 = 1/2$.

This distribution depends on the value of the order parameter $r$. In order to make this solution self-consistent, the value of $r$ has to be determined from the classical Kuramoto results corresponding to an ensemble of oscillators with frequency distribution $G(\omega,kr)$. That is, $r$ is equal to the average of $\exp(i\theta)$ over all the oscillators. As shown, e.g., in Refs.~\cite{kuramotoreview}, the average  over drifting oscillators ($|\omega|> kr$) is zero, and $r$ is thus determined entirely by the locked oscillators ($|\omega| < kr$) whose phase angles are given by Eq.~(\ref{eq5}). Thus we obtain 
$$r = \int_{-kr}^{kr} G(\omega,kr)\sqrt{1-(\omega/kr)^2}d\omega.$$
Besides the solution $r=0$, other possible values of $r$ are given by  the solutions of the nonlinear equation \cite{kuramotoreview}
\begin{equation}\label{usukura}
1 = k\int_{-1}^{1}G(zkr,kr)\sqrt{1-z^2}dz.
\end{equation}

We now study numerically the solutions of Eqs.~(\ref{ge}) and (\ref{usukura}). As we will later argue, noise in the $\theta$ evolution (either extrinsic or due to finite $N$) causes the dynamics to be insensitive to the choice of confining potential $V(\omega)$. Therefore, for simplicity, we will choose an infinite potential well to simplify the analysis. The potential is defined as $V(\omega) = 0$ if $|\omega| < L$, and $V(\omega) = \infty$ otherwise. This corresponds to frequencies that evolve freely in a box of size $2L$ with hard walls. We will later show that $L$ can be chosen large enough so that the dynamics is insensitive to its value. Except for Fig.~\ref{manyls}, all our  plots use $L = 5$.  

Solving Eq.~(\ref{usukura}) numerically, a bifurcating pair of solutions $r_s(k)$ and $r_u(k)$ is found to appear at a finite value of the coupling strength $k=k_1$ as shown in Fig.~\ref{manyls}. The upper branch (solid black line) in these figures was numerically found to be stable, while the lower branch (colored lines in Fig.~\ref{manyls} and dashed line in Figs. \ref{f2}$a$ and \ref{his1}) was numerically found to be unstable.  The trivial solution $r=0$ was numerically found to be stable until the lower branch crosses $r=0$ at a value of the coupling constant $k=k_2$ (see Fig.~\ref{manyls}). For larger coupling strength, $k>k_2$, the nonzero unstable solution disappears and the solution $r=0$ becomes unstable. 

Thus, three regimes are found with our model.  For $k<k_1$, the oscillator ensemble is incoherent, as only the $r=0$ stable solution exists.  For $k>k_2$, only the synchronized state is stable.  These correspond to the traditional regimes of the original Kuramoto model without adaptation \cite{kuramotoreview}.  The third regime corresponds to intermediate coupling strengths on the finite interval $k_1<k<k_2$, where both the synchronized and incoherent states are locally stable, and whose basins of attraction are separated by the unstable solution of Eq.~(\ref{usukura}).  A similar regime was also found in Ref.~\cite{inertia1} for an inertial version of the Kuramoto model without noise [in which $\dot\theta_n$ is replaced by $m \ddot\theta_n + \dot\theta_n$ in (\ref{kuramotoeq})]. In Sec.~\ref{transitions} we address the important issue of noise-induced spontaneous transitions between stable solutions which was not addressed in Ref.~\cite{inertia1}.

\begin{figure}[t]
\centering \epsfig{file =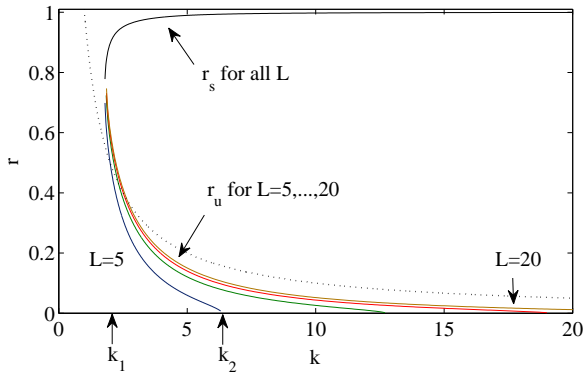, clip =
,width=1.0\linewidth }
\caption{Stable (upper black line, $r_s(k)$) and unstable (lower colored lines, $r_u(k)$) branches are shown for $D=0.01$, $\tau=50$, and $L = 5, 10, 15, 20$  and the curve $kr = \sqrt{2} \sigma$ (dotted line), above which the frequency distribution is normalizable. The values of $k_1$ and $k_2$ for $L=5$ are indicated by vertical arrows at $k_1 \approx 1.8$ and $k_2 \approx 6.37$.} \label{manyls}
\end{figure}

To study the behavior of our model close to the incoherent state $r=0$, we expand $G$ in Eq.~(\ref{ge}) for $(kr/\omega)^2 \ll 1$, finding that the $\omega$ dependence of $G$ for large $|\omega|$ is $G(\omega,kr)\sim (kr/\omega)^{(kr)^2/(2\sigma)^2}$. Thus the frequency distribution $G(\omega)$ can be normalized in $(-\infty,\infty)$ if $kr > \sqrt{2}\sigma$. We conclude that, as long as $kr > \sqrt{2}\sigma$, $G$ should be insensitive to $L$ if the bulk of $G$ is contained within $(-L,L)$.  In contrast, for $kr < \sqrt{2}\sigma$, the frequency distribution is not normalizable in $(-\infty,\infty)$ and thus we expect $G$ to be broadly distributed in $(-L,L)$, and the dynamics to depend on the value of $L$. In particular, the distribution of frequencies for the incoherent state $r=0$ is uniform with $G(\omega) = 1/(2L)$. More generally, $G$ should be insensitive to the specific form of the confining potential $V$ as long as $V$ is negligible in the synchronized state, $V(\sigma) \ll \sigma^2$. This can be interpreted as requiring that $V$ does not itself promote synchronization.




Figure \ref{manyls} shows the stable and unstable branches for various values of $L$ and the curve $kr = \sqrt{2} \sigma$ (dotted line), above which the frequency distribution is normalizable.  Note that the stable solution $r_s(\omega)$ to Eq.~(\ref{usukura}) is above the curve $kr = \sqrt{2} \sigma$ in Fig. \ref{manyls} and is, as expected, insensitive on the value of $L$. One interesting result from Fig. \ref{manyls} is that the upper critical coupling strength $k_2$ depends on the frequency bound $L$.  Recalling that $G(\omega)=1/(2L)$ for the incoherent state, one can integrate Eq.~(\ref{usukura}) to find $k_2 =  4 L/\pi$.   In reality, physical limitations typically bound the natural frequency distribution. However, it is also interesting to consider the unbound case ($V \equiv 0$, or $L = \infty$) which leads to the following scenario:  the oscillators' natural frequencies wander in $(-\infty,\infty)$;  $k_2 = \infty$ and the incoherent state $r=0$ remains stable for all coupling strengths. However, it is important to note that, even though the incoherent state for $L=\infty$ remains stable for arbitrarily large $k$, its basin of attraction shifts to zero  as $k\to \infty$ (the unstable equilibrium $r_u$ approaches zero as $k\to\infty$). The situation is analogous to that applying in the study of the transition of stable laminar pipe flow to turbulence (e.g., Ref.~\cite{eckhardt} and references therein). As in pipe flow, this situation points to the possibly crucial role of noise which we address subsequently.

\begin{figure}[t]
\begin{center}
$\begin{array}{c@{\hspace{.1in}}c@{\hspace{.01in}}c}
a \epsfig{file =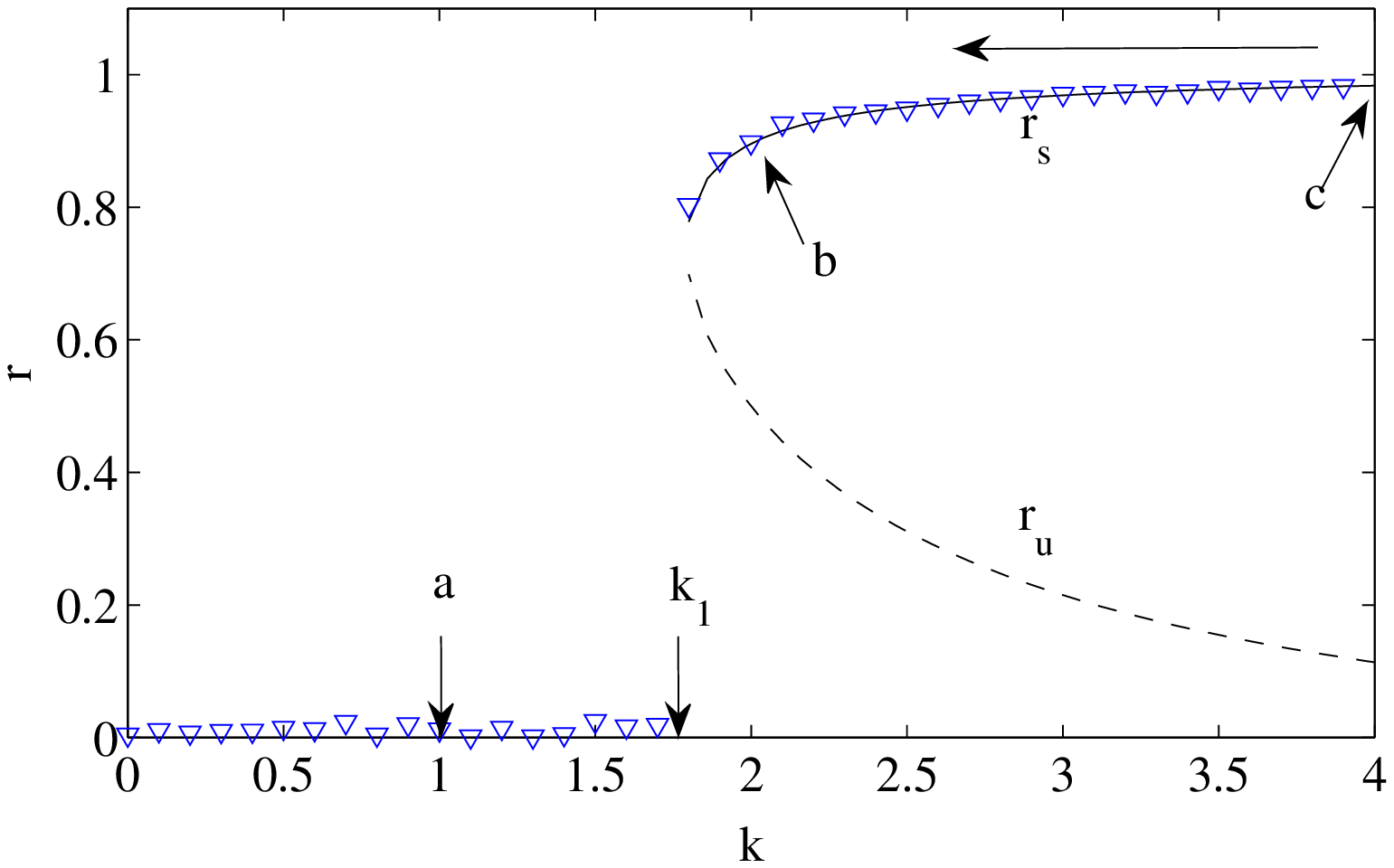, clip =
,width=1.0\linewidth }\\\\
b \epsfig{file =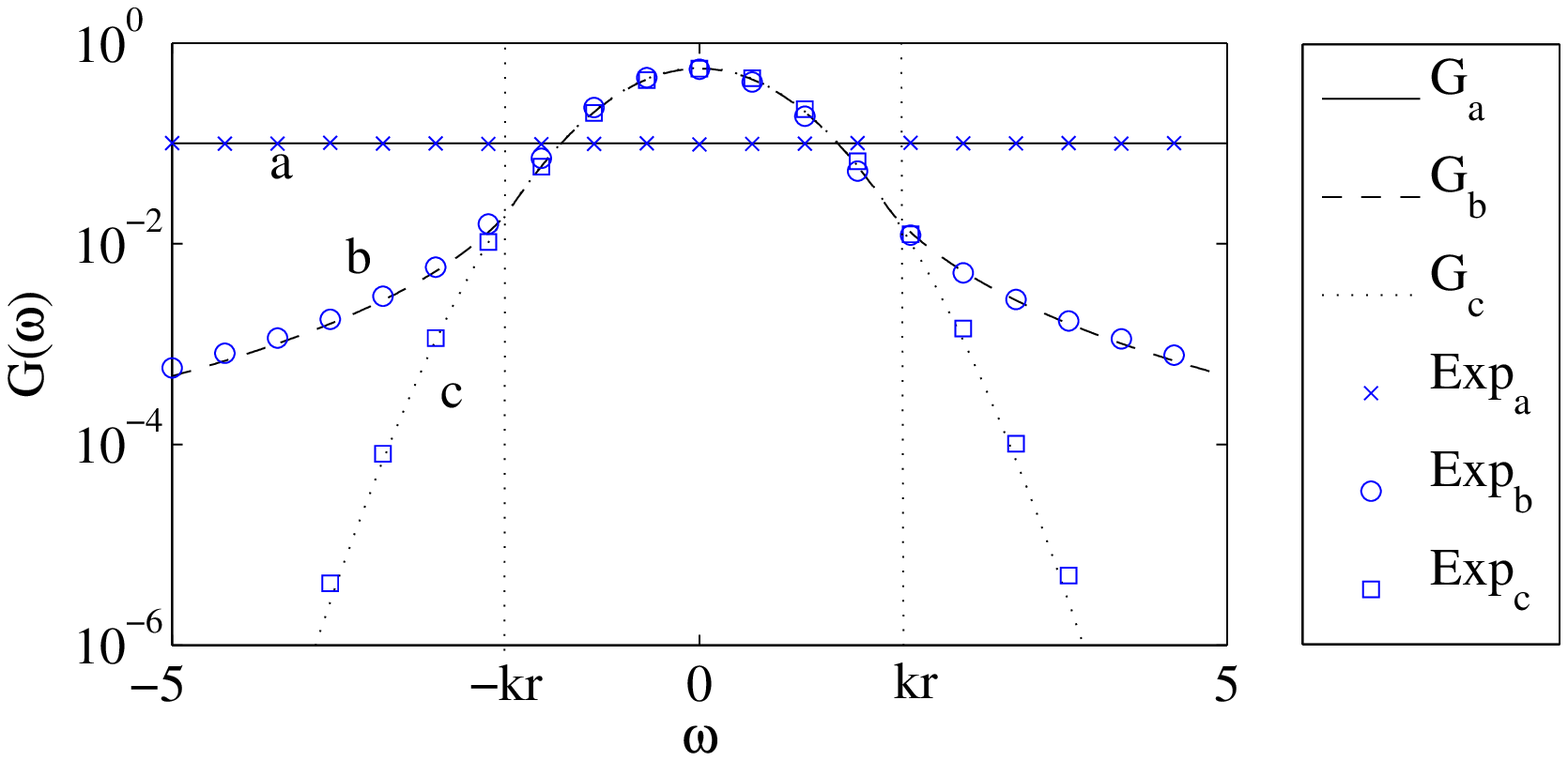, clip =
,width=1\linewidth }
\end{array}$
\end{center}
\caption{$(a)$ Order parameter $r$ obtained from numerical simulation of Eqs.~(\ref{model}) for decreasing values of $k$ for $N = 10^4$ (triangles) with $D=0.01$, $\tau=50$, and $L = 5$. The solid and dashed lines indicate stable $r_{s}(k)$ and unstable $r_{u}(k)$ solutions to Eq. (\ref{usukura}), respectively. The letters a, b and c indicate values of $k$ at which the frequency distribution is sampled for Fig.~\ref{f2}$b$.  $(b)$ Frequency distribution obtained directly from Eqs.~(\ref{model}) (symbols) and from Eq.~(\ref{ge}) (black lines). Curves labeled a, b, and c correspond to $k=1$, $2$, and $4$, indicated by arrows in Fig.~\ref{f2}a. The dashed vertical lines indicate $\pm k r $ for $k = 2$.}
\label{f2}
\end{figure}

%


\subsection{Model Simulation}\label{secmodelb}
In order to test our theoretical results, we compared them with direct simulation of Eqs.~(\ref{model}) using $\tau = 50$ and $D = 0.01$. Due to the stability characteristics of the solutions, hysteresis phenomena and dependence on the initial conditions are expected.  To probe these characteristics, we let $L=5$ and initiate a simulation with strong coupling ($k>k_2$) and with the phases and frequencies of the oscillators clustered  around $\theta_n\approx 0$ and $\omega_n\approx 0$.  The oscillators remain synchronized, and their natural frequencies adopt a distribution given by Eq.~(\ref{ge}). For a given value of $k$, we simulate Eqs.~(\ref{model}) for 1000 seconds and then decrease the value of $k$ by $0.1$, keeping the values of the phases and frequencies (this corresponds to a coarse grained rate $dk/dt \approx 10^{-4}$). As this process is repeated and the value of $k$ decreases below $k_1$, the synchronized solution disappears and the oscillators desynchronize. Figure~\ref{f2}$a$ shows the value of $r$ obtained by this process (triangles). The solid and dashed lines indicate the stable $r_s(k)$ and unstable $r_u(k)$ solutions, respectively obtained from Eq.~(\ref{usukura}). The numerically obtained values of $r$ follow the stable branch found theoretically. 

In Fig.~\ref{f2}$b$ we show the steady-state frequency distribution observed at values of $k$ corresponding to the arrows labeled a, b, and c in Fig.~\ref{f2}$a$. The black solid, dashed, and dotted lines indicate the theoretical expression given by Eq.~(\ref{ge}) normalized on $\omega \in [-5,5]$ for cases a, b, and c.  The cross, circle, and square symbols show the corresponding observed frequency distributions which are in good agreement with the theory. 

To observe hysteresis phenomena similar to that noted in \cite{inertia1}, the system was was brought to steady state with a dispersed frequency distribution described by Eq. (\ref{ge}) for small coupling strength ($k<k_1$).  The coupling strength $k$ was slowly increased until the system underwent an incoherent$\rightarrow$synchronized state transition at the transition coupling strength $k^*$, which is found on the interval $k_1<k^*<k_2$. The precise value of $k^*$ fluctuates slightly from run to run, but its mean is observed to  depend on the ensemble size $N$.  This is shown in Fig.~\ref{his1}, where $k^*$ approaches $k_2$ as $N$ increases, as previously noted in \cite{inertia1} for the inertial Kuramoto model.  This is due to fluctuations in the order parameter $\Delta r \sim N^{-1/2}$ resulting from the system's finite size: it is hypothesized that fluctuations cause the system to cross the barrier imposed by the unstable solution to Eq. (\ref{usukura}) (dashed line in Fig. \ref{his1}).  When the size of these fluctuations becomes large enough to place $r$ above the unstable solution, the oscillators begin to synchronize and the value of the order parameter increases to the value corresponding to the stable solution (upper solid line). 

It should also be noted that, for this simulation with temporally increasing coupling strength, the $k^*$ approach $k_1$ as the simulation duration for each $k$ is increased.  In other words, the hysteretic nature of this system depends not only on the size of the network (as noted in \cite{inertia1}), but also on the rate at which the coupling strength $k$ is varied.  We hypothesize this phenomenology to also describe other Kuramoto-type models with hysteretic behavior (e.g., \cite{inertia1,waishing}).  The fluctuations of the order parameter $r$ are stochastic, and thus the time required for the transition to occur is a random variable.  The longer a simulation is run at constant coupling strength $k_1<k<k_2$, the more likely an incoherent $\rightarrow$ synchronized transition has occurred.  In fact, oscillations between states, as hypothesized in \cite{inertia1}, were observed for our model in this bistable regime (see Fig.~\ref{FFF}).  Describing such spontaneous state transitions is the focus of the next section of our paper.  

\begin{figure}[t]
\centering \epsfig{file =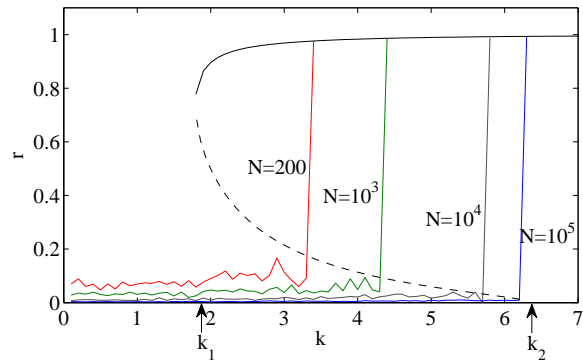, clip =
,width=1.0\linewidth }
\caption{For increasing coupling strength, synchronization occurs for each network when the order parameter fluctuations $\Delta r$ allow $r$ to surmount the barrier of the unstable solution $r_{u}(k)$ (dashed line). Simulation used $D=0.01$, $\tau=50$, and $L = 5$. Note that the transition coupling strengths $k^*$ approach $k_2$ as network size $N$ increases. } \label{his1}
\end{figure}

\section{Spontaneous State Transitions}\label{transitions}
Given the observed phenomenology of fluctuations driving the system from one stable solution to another across an unstable solution, it is natural to conjecture that, for a fixed value of $k$, the average time $\tau_{sync}(N)$ for a transition from the incoherent state $r \approx 0$ to the coherent state $r \sim 1$ can be obtained by treating the problem as an escape over a potential barrier under the influence of random noise (see Fig.~\ref{pics}).  Conceptually, it is helpful to relate such transitions to a Brownian particle moving from one equilibrium to a second by traversing an energy barrier under the influence of random noise. For the case of oscillator system state transitions, fluctuations of the order parameter $\Delta r$ occur due to a network's finite size $N$ and are akin to random noise. In addition, in some applications, Eq.~(\ref{model}) may be subject to extrinsic noise \cite{kuramotoreview}.

For the traditional Kuramoto model, understanding finite size fluctuations $\Delta r$  has been a major area of interest \cite{kuramotoreview, Fluct}.  In general, fluctuations are typically $O(N^{-1/2})$, although it has been shown that these fluctuations increase in amplitude near the critical coupling $k_c$ \cite{Fluct} for a traditional Kuramoto oscillator system.  Similarly, for our model, fluctuations in $r$ were observed to be larger in the bistable regime than in the traditional Kuramoto regimes.  However, as with the traditional Kuramoto model, further study of these fluctuations for our model remains open to future research.

\subsection{State Transition Analysis}
In order to study the statistics of spontaneous synchronization transitions, we will assume that finite-size fluctuations can be described approximately as produced by uncorrelated Gaussian noise acting on the 1-dimensional dynamics of the order parameter.  Treating finite-size fluctuations as an uncorrelated Gaussian noise term has already proven sucessful in studying synchronization of Kuramoto oscillators in networks \cite{Restrepo05}. Consequently, let us assume that the macroscopic dynamics of the order parameter $r$ can be described by a Langevin equation of the form
\begin{equation}\label{LL} 
\dot{r} = -U'(r,k) + L(t),
\end{equation}
where $U(r,k)$ is an unknown pseudo-potential, $U'(r,k)=\partial U/\partial r$, and $L(t)$ is an uncorrelated Gaussian noise term such that $\langle L(t)\rangle=0$ and $\langle L(t)L(t')\rangle=2\Gamma\delta(t-t')$. Since the noise represents finite-size fluctuations, the diffusion coefficient $\Gamma$ will be assumed to be inversely proportional to $N$, or $\Gamma \propto 1/N$. Note that this is consistent with $\Delta r$ being $O(N^{-1/2})$ for the dynamics of $r$ modeled as a linear Ornstein-Uhlenbeck process for the incoherent state with $k<k_1$.
\begin{figure}[t]
\centering \epsfig{file =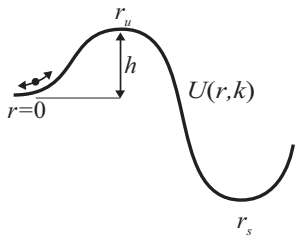, clip =
,width=.45\linewidth }
\caption{State transitions parameterized by $r$ for $k_1<k<k_2$ are schematically shown as Brownian motion in a 1-dimensional energy landscape with two stable equilibria.} \label{pics}
\end{figure}

In the bistable regime, $k_1 < k < k_2$, we assume $U(r,k)$ to be of the form shown in Fig.~\ref{pics}.  Potentials of this type have received much attention in the literature for studying Brownian motion in bistable potentials and for describing chemical reactions.  We will draw on this research and use Kramer's escape time equation \cite{vankampen}, which describes the mean first-passage time $\tau_{esc}$ for a particle subject to random noise with diffusion coefficient $\Gamma$ to escape over a potential barrier of height $h$, and is given by $\log(\tau_{esc}) \varpropto h/\Gamma$. Recalling that $\Gamma \varpropto 1/N$, we conclude that the mean first-passage time (i. e., wait time before synchronization) for our bistable Kuramoto system depends exponentially on $N$, yielding $\tau_{sync} \varpropto e^{KN}$ for some constant $K$.  

A similar analysis can also be done on the regime where the incoherent state is unstable, where we are interested in the average time required for an incoherent system ($r\sim 0$) to synchronize. To first order, the dynamics for small $r$ is described by $\dot{r}= \alpha r + L(t)$, with $\alpha$ being a positive constant. Taking $r(0)=0$ and setting $\langle r(t^*)^2\rangle \equiv {r^*}^2$, we can estimate for large $N$ the time $t^*$ it takes for the order parameter to reach a given threshold $r=r^* \gg \sqrt{\Gamma/\alpha}$ as $t^* \sim \log{\Gamma^{-1}} \sim \log{N}$.
Thus, the waiting time $\tau_{sync}$ grows logarithmically with $N$ in the strong coupling regime ($k>k_c$ in the Kuramoto model or $k>k_2$ in our model). 

Although this paper focuses on the model described by Eqs.~(\ref{model}), the above estimates may apply to other Kuramoto-type models \cite{inertia1, inertia2,waishing}.

\subsection{State Transition Simulation}
To test the previous findings, statistics were compiled for our adaptive Kuramoto system by simulating 100 realizations of synchronization for an initially incoherent system.  For each realization, at a constant coupling strength $k$ the initial natural frequencies and phases were chosen randomly ($\theta_n$ uniform in $[0,2\pi)$, and $\omega_n$ uniform in $[-5,5]$).  Once the order parameter exceeded a given threshold $r^*$ ensuring synchronization had occurred, the time before synchronization was recorded and simulation stopped.

 \begin{figure}[t]
\centering \epsfig{file =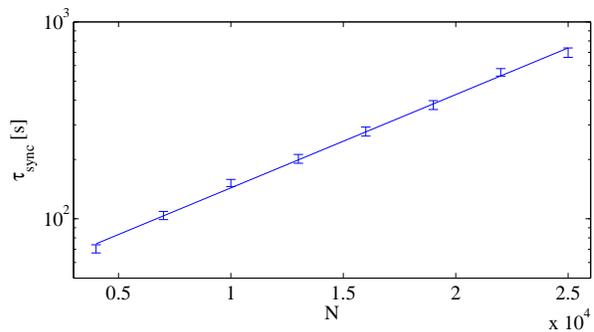, clip =
,width=1\linewidth }
\caption{Synchronization time $\tau_{sync}$ averaged over $100$ realizations as a function of the number of oscillators $N$ for $k=6$, which is within the bistable regime. ($D=0.01, \tau=50,$ and $L=5$)} \label{conje}
\end{figure}

Statistics of incoherence$\to$synchronization transitions for the bistable regime are shown in Fig.~\ref{conje}, where $\log(\tau_{sync}(N))$ vs. $N$ is plotted for $k = 6$. $\tau_{sync}$ is defined as the average time required for the order parameter to first reach $r^*=0.7$. In principle any coupling strength $k$ within the bistable regime could be used; however, to decrease simulation time $k$ was chosen to be close to $k_2\approx6.37$. Error bars are included to show statistical uncertainty.  As the plot shows, $\log(\tau_{sync}(N))$ is well described by a straight line, which is consistent with the supposition that the transition times can be described by Kramer's escape time formula. 

For comparison, $\tau_{sync}$ is shown in Fig. \ref{conje2} for synchronization with the incoherent state being unstable ($k > k_2$).  From this figure we confirm that $\tau_{sync} \varpropto \log{N}$, which is consistent with unstable exponential growth of perturbations from the $r=0$ incoherent state. 

\begin{figure}[t]
\centering \epsfig{file = 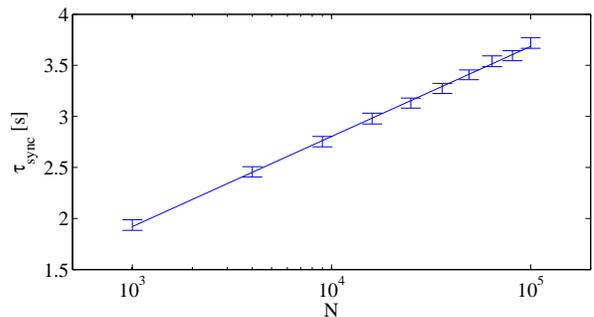, clip =
,width=1\linewidth }
\caption{Synchronization time $\tau_{sync}$ averaged over $100$ realizations as a function of the number of oscillators $N$ for $k=7>k_2$. ($D=0.01, \tau=50,$ and $L=5$). Note that the scale is different than that of Fig.~\ref{conje}.} \label{conje2}
\end{figure} 

Figure \ref{FFF} shows fluctuations between the synchronized and incoherent states for a case where the coupling strength is within the bistable range.  Note that since transitions between states are related to the height $h$ of the pseudo-potential barrier relative to each respective equilibrium (see Fig.~\ref{pics}), fluctuations between states can only be observed when the barrier heights are roughly equal and when the system is observed for a duration in which transition-events should occur.  For example, if the barrier height is large and the finite system is large (large $N$), the order parameter $r$ will undergo small fluctuations and state transitions would be rare.  At the same time, if the barrier height is much larger for a particular state, then the system will remain in that state for the majority of time and transitioning out of that state would also be rare.  For the model parameters chosen in our simulation, we found that spontaneous bidirectional transitions could only be observed for small numbers of oscillators ($N=10$ in Fig.~\ref{FFF}) and for coupling strengths in the bistable regime just above $k_1$ (below which the coherent solution disappears). In general, for $k_1<k<k_2$, we find that synchronized$\to$incoherent transitions are very rare, implying that the barrier height for the synchronized state is generally larger than the barrier height for the coherent state (as shown schematically in Fig~\ref{pics}).  

\begin{figure}[t]
\centering \epsfig{file = 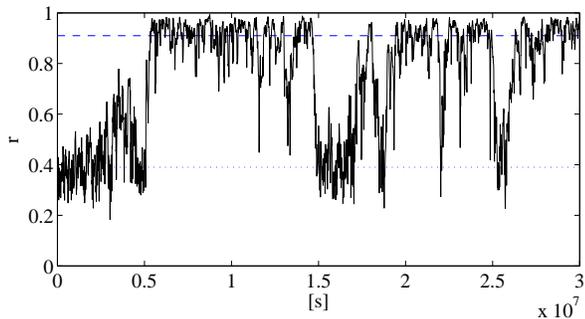, clip =
,width=1\linewidth }
\caption{Spontaneous bidirectional transitions between the synchronized (dashed) and incoherent (dotted) states are observed for $N=10$, $k=1.9$, $D=0.01, \tau=50,$ and $L=5$.  Note that because of the small system size, the incoherent state has an average order parameter of $\langle r \rangle \sim0.4$. } \label{FFF}
\end{figure}

\section{Discussion}\label{secdiscussion}
Our results discussed above are in striking agreement with observations of rhythmically clapping audiences \cite{clapping, Xenides}.  In particular, as opposed to the behavior of the classical Kuramoto model without adaption, the transition to synchronized clapping occurs after a relatively long waiting time, and once it starts the order parameter quickly achieves its steady state. Previous models of this phenomenon have artificially altered the frequency distribution \cite{clapping} or introduced additional dynamics such as a time-dependent tendency of the oscillators to synchronize \cite{Xenides}. In contrast, the long waiting times arise in our model as a natural consequence of the dynamics. Although we have found that all-to-all coupling leads to waiting times that depend exponentially on the number of oscillators, shorter waiting times are expected for local coupling such as that describing clapping synchronization in a large venue. 


Another possible application of our model is circadian rhythms \cite{drift}, which have been modeled by ensembles of  Kuramoto oscillators with drifting, nonadaptive frequencies \cite{Rougemont}. Because of the importance of synchronization in this system, evolutionary pressures might have led to frequency adaptation. Our model generalizes previous models \cite{Rougemont} by allowing for frequency adaptation. By removing frequency coupling (i.e., $\tau\to\infty$) and assuming a quadratic form for the potential $V(\omega)$, our model [Eqs.~(\ref{model})] recovers the model of coupled circadian oscillators presented in \cite{Rougemont}.


Our results are somewhat related to the Kuramoto model with inertia (Eq.~(\ref{kuramotoeq}) with $\dot\theta_n$ replaced by $m \ddot\theta_n + \dot\theta_n$ \cite{ermentrout,inertia1,inertia2}), which is equivalent to 
\begin{align}\label{ine}
\dot{\theta}_n &= \omega_n,\\
\dot{\omega}_n &= \tau^{-1}\left[-dV_n(\omega_n)/d\omega_n + \frac{k}{N} \sum_{m=1}^N\sin(\theta_m-\theta_n) +\eta_n\right],\nonumber
\end{align}
where $V_n(\omega_n) = \frac{1}{2}(\omega_n-\Omega_n)^2$, with $\Omega_n$ constant for each oscillator.
However, the differences between this model and our model are significant. First, in contrast to (\ref{ine}), our model couples both phases and frequencies.
Second, as a consequence of our two types of coupling, we are able to introduce two time scales, with the frequency adaptive time scale being slower than that of the phase dynamics. We believe that this two time scale dynamics will be crucial to the modeling of the various potential applications mentioned in Sec.~\ref{introd} (e.g., clapping audiences). [Note that simulations were conducted to investigate the effect of closing the timescale gap. By keeping $\sigma$ constant and reducing $\tau$ (i.e., by also increasing $D$), it was found that no qualitative differences were observed as long as $\tau>5$.]


The analysis presented here to describe fluctuation-induced spontaneous transitions from incoherence to synchronization for our adaptive model could also be applicable to other Kuramoto-type systems with hysteretic behavior. Such systems include Kuramoto models with an added inertial term \cite{inertia1,inertia2} and situations where there is a heterogeneous distribution of interaction time delays \cite{waishing}. Various questions remain to fully understand the dynamics of the observed transitions. While order parameter fluctuations are typically $O(N^{-1/2})$, this is not always the case and a better understanding of these fluctuations is needed.  Similarly, the existence of a pseudo-potential $U(r,k)$ was assumed [Eq.~\ref{LL}], but its shape and dependence on $k$ remain to be investigated.

\section{Conclusions}\label{secconclusions}
We have presented a new model to study the synchronization of Kuramoto oscillators that are able to slowly adapt their natural frequencies to promote synchronization, but are inhibited from doing so completely by the influence of noise. We found that the interplay of noise and adaptation results in bistability and hysteresis. In the bistable regime, finite size effects induce incoherent$\to$synchronized state transitions (or, when $N$ is small, vice versa), which are well described as a 1-dimensional Kramer escape process on the order parameter $r$.  For an oscillator ensemble governed by our adaptive model with all-to-all coupling, it was shown that the time $\tau_{sync}$ required for the system's state to transition from incoherent to synchronized depended exponentially on $N$ in the bistable regime ($k_1 < k < k_2$) and logarithmically for strong coupling ($k>k_2$).

To our knowledge, this work is the first to analyze spontaneous synchronization at constant coupling strength as a $1$-dimensional stochastic escape process.    It is expected that the analysis presented in this paper is also valid for other Kuramoto-type models with hysteretic behavior \cite{inertia1,inertia2,waishing}.   

The work of D. Taylor and J. G. Restrepo was supported by NSF (Applied Mathematics) and the work of E. Ott was supported by the NSF (Physics) and by the ONR (N00014-07-0734).

\bibliographystyle{plain}

\begin{thebibliography}{99}

\bibitem{dano} S. Dano, M. F. Madsen, and P. G. Sorensen, Proc. Natl. Acad. Sci. USA {\bf 104}, 12732 (2007).

\bibitem{millenium} S. H. Strogatz, D. M. Abrams, A. McRobie, B. Eckhardt, and E. Ott, Nature {\bf 438}, 43 (2005).

\bibitem{josephson} K. Wiesenfeld, P. Colet, and S. H. Strogatz, Phys. Rev. Lett. {\bf 76}, 404 (1996).

\bibitem{Fillatrella} G. Fillatrella, A. H. Nielsen, and N. F. Pedersen, Euro. Phys. J. B {\bf61}, 485-491 (2008).

\bibitem{Cross} M. C. Cross, A. Zumdieck, R. Lifshitz, and J. L. Rogers, Phys. Rev. Lett. {\bf 93}, 224101 (2004).

\bibitem{ermentrout} B. Ermentrout, J.  Math. Bio. {\bf 29}, 571 (1991). 

\bibitem{kuramoto} Y. Kuramoto, {\it Chemical Oscillations, Waves, and Turbulence}, Springer (1984).

\bibitem{kuramotoreview} S. H. Strogatz, Physica D {\bf 143}, 1-20 (2000);  E. Ott, {\it Chaos in Dynamical Systems}, Cambridge University Press (2002), Sec.~6.5;  J. A. Acebr\'on, L. L. Bonilla, C. J. P. Vicente, F. Ritort, and R. Spigler, Reviews of Modern Physics, {\bf 77}, 137 (2005).

\bibitem{otts} E. Ott and T. M. Antonsen, Chaos {\bf 18}, 037113 (2008); {\bf 19}, 023117 (2009).

\bibitem{adapta} P. Seliger, S. C. Young, and L. S. Tsimring, Phys. Rev. E, {\bf 65}, 041906 (2002); P. M. Gleiser and D. H. Zanette, Euro. Phys. J. 53 {\bf 2} 233-238 (2006); C. Zhou and J. Kurths, Phys. Rev. Lett. {\bf96}, 164102 (2006); Y. L. Maistrenko, B. Lysyansky, C. Hauptmann, O. Burylko, and P. A. Tass, Phys. Rev. E, {\bf 75}, 066207 (2007); Q. S. Ren and J. Y. Zhao, Phys. Rev. E {\bf 76}, 016207 (2007). G. He and J. Yang, Chaos 35 {\bf5}, 1254-1259 (2007); T. Aoki and T. Aoyagi, Phys. Rev. Lett. {\bf 102}, 034101 (2009).

\bibitem{clapping} Z. N\'eda, E. Ravasz, Y. Brechet, T. Vicsek, and A. L. Barab\'asi, Nature {\bf 403}, 849 (2000); Z. N\'eda, E. Ravasz, T. Vicsek, Y. Brechet,  and A. L. Barab\'asi, Phys. Rev. E {\bf 61}, 6987 (2000).

\bibitem{Xenides} D. Xenides, D. S. Vlachos, and T. E. Simos, J. Stat. Mech.  P07017 (2008).

\bibitem{drift} E. Nagoshi, C. Saini, C. Bauer, F. Naef, and U. Schibler, Cell {\bf 119}, 693 (2004); D. K. Welsh, S. H. Yoo, A. C. Liu, J. S. Takahashi, and S.A. Kay, Curr. Biol. {\bf 14}, 2289 (2004); A. J. Carr and D. Whitmore, Nat. Cell Biol. {\bf 7}, 319 (2005).


\bibitem{inertia1} H. A. Tanaka, A. J. Lichtenberg, and S. Oishi, Physica D {\bf 100}, 279 (1997).

\bibitem{inertia2} J. A. Acebr\'on and R. Spigler, Phys. Rev. Lett. {\bf 81}, 2229 (1998); J. A. Acebr\'on, L. L. Bonilla, and R. Spigler,  Phys. Rev. E {\bf 62}, 3437 (2000).

\bibitem{Rougemont} J. Rougemont, Phys. Rev. E {\bf 73}, 011104 (2006); J. Rougemont and F. Naef, Mol. Sys. Biol. {\bf 3}, 93 (2007).

\bibitem{eckhardt} B. Eckhardt, Phil. Trans. Roy. Soc. A {\bf 367}, 449 (2009).

\bibitem{waishing} W. S. Lee, E. Ott, and T. M. Antonsen, Phys. Rev. Lett. {\bf 103}, 044101 (2009).

\bibitem{Fluct} H. Daido, J. Phys. A: Math. Gen. {\bf 20}, L629-L636 (1987), Prog. Theor. Phys. {\bf 81}, 727-31 (1989), Prog. Theor. Phys. Suppl. {\bf 99}, 288-294 (1989), J. Stat. Phys. {\bf 60}, 753 (1989); M. A. Buice and C. C. Chow, Phys. Rev. E 76, 031118 (2007).

\bibitem{Restrepo05} J. G. Restrepo, B. R. Hunt, and E. Ott, Phys. Rev. E {\bf 71}, 036151 (2005).

\bibitem{vankampen} N. G. Van Kampen, {\it Stochastic Processes in Physics and Chemistry}, Elsevier (2007).



\end{thebibliography}

\end{document}